Hiring Fairly in the Age of Algorithms

Allan Costa

Chris Cheung

Max Langenkamp

{acosta, ccheung9, maxnz}@mit.edu

## Table of Contents

**Abstract**

Widespread developments in automation have reduced the need for human input. However, despite the increased power of machine learning, in many contexts these programs make decisions that are problematic. Biases within data and opaque models have amplified human prejudices, giving rise to such tools as Amazon's (now defunct) experimental hiring algorithm, which was found to consistently downgrade resumes when the word "women's" was added before an activity.

This article critically surveys the existing legal and technological landscape surrounding algorithmic hiring. We argue that the negative impact of hiring algorithms can be mitigated by greater transparency from the employers to the public, which would enable civil advocate groups to hold employers accountable, as well as allow the U.S. Department of Justice to litigate. Our main contribution is a framework for automated hiring transparency, algorithmic transparency reports, which employers using automated hiring software would be required to publish by law. We also explain how existing regulations in employment and trade secret law can be extended by the Equal Employment Opportunity Commission and Congress to accommodate these reports.

## I.    Introduction

Several companies have begun to promise and use software that can instantly evaluate thousands of job applicants on the basis of simple features, such as their resumes or cover letters[1]. They promise not only tremendous scaling, but objective, 'anti-biased' evaluation[2] of candidates. In reality, however, these algorithms often reflect or even amplify the prejudices that have historically affected humans[3], yet due to a lack of transparency about how the algorithms are created and used, which we term the 'transparency gap', candidates are unable to demonstrate employment discrimination.

---

[1] See https://www.allyo.com/, https://modernhire.com/, https://www.HireVue.com/, https://harver.com/ , https://learn-about.hiring.monster.com/Hiring-Solutions/smart-hiring?intcid=RC_Body_Context_GEN, https://business.linkedin.com/talent-solutions#

[2] Ajunwa, Ifeoma, *The Paradox of Automation as Anti-Bias Intervention* (March 10, 2016). Ifeoma Ajunwa, The Paradox of Automation as Anti-Bias Intervention, 41 Cardozo, L. Available at SSRN: https://ssrn.com/abstract=2746078 or http://dx.doi.org/10.2139/ssrn.2746078

[3] Miller, Claire Cain. "Algorithms and bias: Q. and A. with Cynthia Dwork." *New York Times* 11 (2015).

The three key stakeholders within this article are the *candidates* (people seeking employment), *hiring software companies* (companies such as HireVue[4] which produce and sell the automated hiring software), and *employers* (large companies, such as Unilever, which use the automated hiring software).

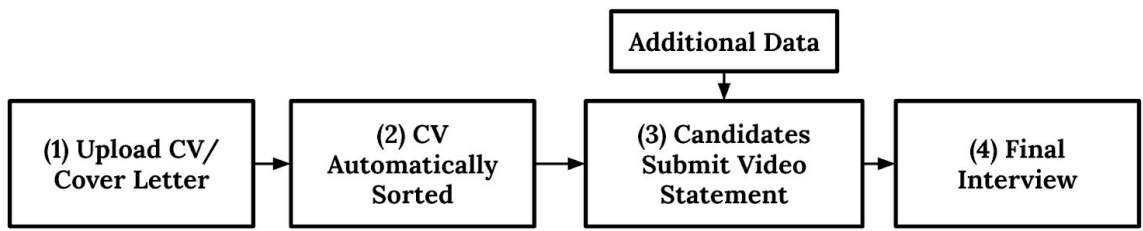

**Figure 1: Stages of Candidate Screening**

In this article, we focus on the early stages of candidate screening. As visible in Figure 1, there are four stages to the automated hiring process, beginning with the resume submission, and ending with an offer. We are concerned specifically with stages 2 and 3 of the automated hiring process. Many of the biggest hiring companies sell services that either allow a sorting of candidates by their resumes[5], or through an interview assessment process. Any significant solution to the problems posed by hiring thus needs to address these stages of the hiring process.

Part II discusses the implications of hiring algorithms; their potential as an unbiased assessment, and how they currently fail in practice due to a lack of transparency. Part III lays out the technical foundation of algorithmic fairness, and discusses four ways in which algorithms may be misused. Part IV argues that the current hiring law inadequately addresses automated hiring bias, and that trade secret law is used by software companies as a shield to deflect accountability. Finally, Part V proposes *algorithmic transparency reports*, which modifies existing legislation to address the current 'transparency gap' that has hitherto allowed hiring software companies and employers to escape responsibility.

---

[4] https://www.HireVue.com/

[5] Indeed 2016b. Indeed Hits Record 200 Million UniqueVisitors. Indeed.(2016).http://blog.indeed.com/2016/02/08/indeed-200-million-unique-visitors/, CareerBuilder 2016. Careerbuilder About Us. Careerbuilder. (2016).http://www.careerbuilder.com/share/aboutus/, LinkedIn 2018, Recruiting Solutions (2019) https://business.linkedin.com/talent-solutions

## II.    Implications of Automated Hiring

In this section, we explore past and present hiring discrimination, and then discuss the promise of automated hiring. We then investigate the pitfalls of algorithmic hiring assessments, and how a lack of transparency exacerbates the problem.

### A.  Bias in Employment

The corporate hiring process of both the past and present is rife with bias. By bias, we mean a pattern of decision-making that arbitrarily favors a certain subgroup over others[6]. This is in line with the common accepted use of the word[7]. Many humans demonstrate some degree of cognitive bias[8], and this bias influences hiring decisions. If a bias is explicitly against a racial group, as we observed prior to the Civil Rights Act of 1962[9], before which it was legal to deny jobs to African Americans on the basis of their race, then we consider such bias to be unacceptable. Instead, if people hire their children to do chores around the house instead of giving everyone in the community an opportunity, this type of bias is seen as acceptable.

However, many human biases, once exposed, tend to be undesirable. One example of such a bias is the bandwagon effect — many coworkers believe one candidate is fit for the job, but another coworker may see inherent flaws in that candidate. Instead of voicing his or her opinion about these flaws, the coworker chooses to bury his or her ideas in order to avoid irking the other coworkers[10]. Another example of bias is stereotyping, where a belief about a group of people is applied to every instance of its members. This phenomenon has had uneven effects on different demographic subgroups[11]. A set of 476 hiring audits in Washington D.C. and Chicago conducted in 1990 found that this stereotype bias negatively affected black job seekers in particular[12]. The study found that one in five audits involved a white candidate who proceeded further in the interview process than their black counterpart of equal qualifications, and only 5% of audits had a black candidate receive a job when an equally qualified white candidate did not. Whether it is a bias towards physically attractive people[13], a racial bias[14], or even the tendency to underestimate one's own bias[15], humans systematically make decisions that result in seemingly unfair outcomes for members of different groups.

---

[6] Note: Bias in humans is generally referred to as 'cognitive bias', while bias in programs is generally referred to as 'algorithmic bias'.

[7] Dickson, Ben, et al. "What Is Algorithmic Bias?" *TechTalks*, 27 Mar. 2018, https://bdtechtalks.com/2018/03/26/racist-sexist-ai-deep-learning-algorithms/.

[8] Moss-Racusin, Corinne A., et al. "Science faculty's subtle gender biases favor male students." *Proceedings of the National Academy of Sciences* 109.41 (2012): 16474-16479. Vedantam, Shankar. "See no bias." *Washington Post* 21 (2005).

[9] "Jim Crow Laws." *PBS*, Public Broadcasting Service, https://www.pbs.org/wgbh/americanexperience/features/freedom-riders-jim-crow-laws/.

[10] "8 Types of Hiring Bias That Exclude the Candidates You Need | Hire by Google." *Google*, Google, https://hire.google.com/articles/hiring-bias/.

[11] González, M José, et al. "The Role of Gender Stereotypes in Hiring: A Field Experiment." *European Sociological Review*, vol. 35, no. 2, 2019, pp. 187–204., doi:10.1093/esr/jcy055.

The purpose of this article is not to argue for precisely which types of bias should be considered acceptable, although we do discuss algorithmic bias in detail in Section III, but to draw attention to the multifaceted nature of bias, and argue that transparency is necessary if we are to understand and redress these different types of biases.

### B. The Promise of Expedient Neutrality

In order to eliminate the factor of "human" bias in hiring and speed up the process of screening, employers have implemented the use of digital hiring tools and algorithms for the past twenty years[16]. The late-1990s witnessed the birth of online job listings, such as Monster and Craigslist, which expedited the job-search experience for both employers and job candidates[17]. In the mid-2000s, the emergence of LinkedIn allowed for any employer to easily and efficiently search for a candidate because candidate profile data was brought online and made freely available[18].

Today, there is real promise that hiring algorithms can help mitigate recruiters' cognitive biases. Kuncel et al. argue in their article *In Hiring, Algorithms Beat Instinct*[19] that algorithms provide more objective outcomes than humans, by discarding less relevant information such as "applicants' complements or remarks on arbitrary topics". The hiring software company Pymetrics reports that companies increased diversity by 20-100% when using their software[20]. Of course, other hiring software companies have used similar arguments to contend that their algorithms eliminate the biases of human recruiters[21].

---

[12] Turner, Margery Austin, et al. *Opportunities Denied, Opportunities Diminished: Discrimination in Hiring.* Urban Institute Press, 1991.
[13] Shahani-Denning, Comila. "Physical attractiveness bias in hiring: What is beautiful is good." *Hofstra Horizon* (2003): 14-17.
[14] Sporer, Siegfried Ludwig. "The cross-race effect: Beyond recognition of faces in the laboratory." *Psychology, Public Policy, and Law* 7.1 (2001): 170.
[15] Pronin, Emily. "Perception and misperception of bias in human judgment." *Trends in cognitive sciences* 11.1 (2007): 37-43.
[16] "An Examination of Hiring Algorithms, Equity, and Bias." *Help Wanted*, https://www.upturn.org/reports/2018/hiring-algorithms/.
[17] Overell, Michael. "The History of Innovation in Recruitment Technology and Services." *TechCrunch*, TechCrunch, 29 Oct. 2016, https://techcrunch.com/2016/10/29/the-history-of-innovation-in-recruitment-technology-and-services/.
[18] Ibid. (see 16)
[19] Nathan R. Kuncel, Deniz S. Ones & David M. Klieger, In Hiring, Algorithms Beat Instinct,

Aside from the promise of neutrality, these algorithms save companies a significant amount of time. Hilton, for instance, claims to have reduced the hiring process from six weeks to five days by using the hiring software company HireVue[22]. For context, before implementing HireVue's models, Hilton used an hour-long questionnaire per candidate that often went uncompleted, meaning using HireVue's new technology increased the rate of completion and decreased overall time taken in their hiring process. As we will discuss in Section III, many of these software tools operate using machine learning, in which a 'training set' containing good and bad examples (e.g. past hired and rejected candidates) is used to learn how to make decisions. Within automated hiring, as established in the introduction, the use cases we are most concerned with are *resume parsing algorithms*, which are able to take in a candidate's resume and automatically calculate the 'employability' of a candidate, and *video statement analysis algorithms*, which take in a video of a candidate interview to calculate 'employability'.

While these algorithms have the potential to positively change the hiring process, they currently serve to amplify biases of the past, and threaten to unfairly violate candidates' privacy.

### C. The Unfortunate Reality

Amazon's attempt at creating a hiring algorithm provides a compelling example. In 2014, a team at Amazon created an algorithm to parse resumes and infer the best candidates using its own workforce over the past 10 years as training data. However, a large majority of its existing workforce was white and male[23] and accordingly the algorithm was systematically biased against female applicants[24]. In relying on the historical training data, the model associated non-white and female individuals with a lack of success. Thus, whenever the algorithm parsed a candidate's resume and found words such as "female" or "women's college", it gave these individuals a lower score and was more likely to decline these candidates a job.

Joy Buolamwini's experiment[25] presents another example. After experimenting with different popular facial recognition programs, she noticed that while her white friends could be classified with high accuracy, Buolamwini's darker-skinned face was not recognized by the model. She observed the same for many of her black friends, and in a follow up study[26] found significantly lower facial recognition accuracy for minorities.

In both Buolamwini's experiment and Amazon's attempted hiring algorithm, algorithmic bias was only exposed because of concerted attempts by third parties to make this information public. In the case of Amazon's hiring, it was the news service Reuters which managed to uncover the project[27], presumably using a secret source within Amazon. Similarly, it was only because Buolomwini had access to the facial recognition engines of technology companies Microsoft, IBM, and Face++ that she was able to expose the racial bias within the algorithm. Buolomwini would likely not be able to perform the same analyses of the algorithms of hiring software companies today. Of the most popular previously mentioned hiring software companies[28], none provided either public access to their algorithms or concrete information about how their algorithms work to mitigate bias. In other words, despite the evidence that today's algorithms are biased, there is a current transparency gap around hiring software that prevents researchers and reporters from exposing algorithmic bias.

### III.     Technical Background

The term 'hiring algorithms' describes a large collection of mathematical tools able to evaluate a candidate's fit for a certain job position, providing a score that is used as input for further hiring decisions. Although 'algorithms' refers to a much broader range of diverse problem-solving methods, in the context of automated hiring they mostly include supervised machine learning models[29], and so we will use the terms 'algorithms' and 'machine learning models' interchangeably.

In technical terms, supervised machine learning consists of a collection of algorithms able to adapt their decision-making process based on exposure to prior data. In other words, the algorithm 'learns' how to make better decisions or draw better conclusions as it is exposed to previously known and labeled information[30], a process called *training*. Numerous methods are then used to ensure that training results in a model able to generalize to new data, so that the algorithm is able to use learned patterns to infer missing properties of new incoming data (see Figure 2). In large datasets, this process is called data mining, the process of finding patterns and producing statistical knowledge out of large datasets.

---

[29] For an example of what we believe are common methods used in automated hiring, see Naim, Iftekhar, et al. "Automated prediction and analysis of job interview performance: The role of what you say and how you say it." *2015 11th IEEE International Conference and Workshops on Automatic Face and Gesture Recognition (FG)*. Vol. 1. IEEE, 2015., and Chuang, Zhang, et al. "Resume parser: Semi-structured chinese document analysis." *2009 WRI World Congress on Computer Science and Information Engineering*. Vol. 5. IEEE, 2009.

[30] In a supervised machine learning setting. See

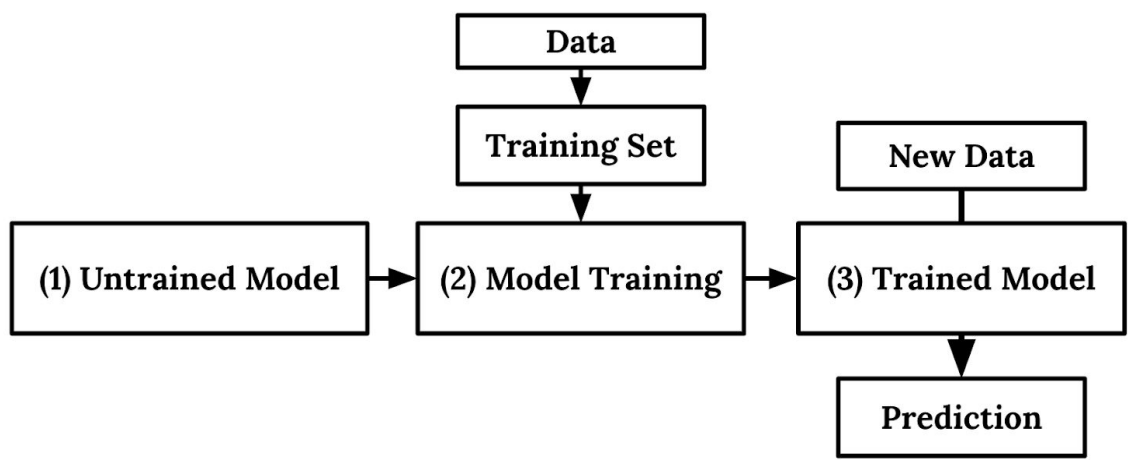

**Figure 2: Machine Learning Model Creation Process**
(1) An algorithm that is able to adapt its inner logic and 'learn' from exposure to labeled data is specified. (2) Labeled data is fed as a training set to the model, which then finds complex correlations on the information provided. (3) The model is then able to infer the label of unlabeled data by extrapolating from previous observations.

In Section II.C, we described several cases where algorithms have unfairly treated subgroups of the population. In this section we will categorize the different factors that typically lead to this unfair treatment.

### A. When Algorithms Go Wrong

Let us now introduce an example for the purposes of illustration: the Acme Hiring Corporation (Acme) is a hiring software company that intends to "revolutionize the hiring process by eliminating human bias with AI". Acme's technology can parse resumes and extract relevant information such as names, addresses, or lists of past experiences. Another product can analyze interview videos, and supposedly measure complex personality traits like discipline or resourcefulness, and emotional cues such as excitement or anxiety. The company's main product, the AcmeSuite, is a full decision-making pipeline in which information extracted from resumes, video and job applications can be fed into a final scoring machine learning model.

After years of using Acme's technology for hiring, some clients reported a gradual increase of its white male population, and a corresponding decrease in its minority population. Acme responded to the companies' queries by arguing that it has a specialized group of scientists and psychologists that carefully ensure that algorithmic bias is mitigated in the AcmeSuite software. However, Acme never released any data regarding the performance of the model or any specific means it used to mitigate bias, arguing that such information is proprietary. Despite Acme's effort to ensure fairness in its algorithmic framework, the results from its client provided evidence that some form of discrimination against minorities was taking place; that likely the algorithms were biased.

What could have led to this outcome?

In this paper, we propose four main categories of mistakes leading to biased decisions in the context of algorithmic hiring. They can be summarized as failures in the **Dataset**, **Model Structure**, **Metrics** and of **Application**.

### A.1 Dataset

Because most machine learning models are built by finding statistical patterns in observed data, their predictive power is limited by the correlations in the training dataset. For example, Acme's resume parser might draw conclusions about people based on their name. Having been exposed to enough information about people, such a model will infer that the new data point *Maria* is likely to correspond to a Latina woman, whereas *Joe* might be expected to be the name of a white man. These conclusions will not necessarily correspond to reality, but rather to the statistical correlations found in the training data fed into it. Importantly, modern machine learning is unable to differentiate correlation from causation[31]. This is especially problematic when protected characteristics – such as race – are strongly correlated with negative outcomes — such as unemployment — in part because it does not allow for social change to happen. If, for instance, the data were gathered during a period when it was socially frowned upon for a woman to work, the algorithms trained on this data would replicate the sexist discrimination of that era. Because these patterns are fully encoded in the training information, proper dataset usage is critical to make sure that these models are deciding fairly.

There are several ways in which Acme's dataset could have resulted in a biased machine learning model. It is worth noting that although many of these categories are problematic, bias in machine learning is not always undesirable. For instance, if we want to account for a subgroup that has been historically marginalized, it may be necessary to have a bias *in favor* of the marginalized subgroup.

---

[31] For a discussion about machine learning and causation, see Obermeyer, Ziad, and Ezekiel J. Emanuel. "Predicting the future—big data, machine learning, and clinical medicine." *The New England journal of medicine* 375.13 (2016): 1216. : "machine learning does not solve any of the fundamental problems of causal inference in observational data sets"

Here, we use the classifications proposed by Pauline T. Kim in her studies of algorithm-based discrimination at work, which divides bias into four different categories[32]: *Record Error Bias, Intentional Bias, Statistical Bias, and Structural Bias.*

*Record Error Bias* can occur if Acme's training dataset is incomplete, mistyped or simply has incorrect data. These errors are particularly present in names with uncommon spellings[33] or women who have changed their names. This leads to lower accuracy for underrepresented subgroups[34].

*Intentional Bias* can occur if Acme intentionally encodes characteristics such as race and gender into the model, whether to explicitly discriminate against minorities, or to attempt to counter structural inequalities.

*Statistical Bias* can occur if Acme uses unrepresentative datasets to train its algorithms. The best example of such a bias is the Amazon case[35] mentioned in Section II.C. We saw that the overrepresentation of white men in Amazon's existing workforce trained the algorithm to associate being a white man with being successful, and did not account for the complex historical and sociopolitical picture that led to fewer opportunities for other minorities.

Statistical Bias is especially tricky because the complex and obscure correlations drawn by machine learning models could mean that arbitrary features are used for important decisions[36]. For instance, it could be the case that, in Acme's video training dataset, people who sneezed also performed better - this correlation could easily be highlighted by a statistical model, despite being obviously senseless for any human. In any case, the resulting algorithm would yield a higher score for people who happened to sneeze during the interviews, discriminating against those who didn't.

Finally, even if Acme's dataset was representative and free of errors, it could be the case that such dataset encoded legitimate correlations that, ultimately, can still result in discrimination. This is an example of what Pauline T. Kim calls *Structural Bias*. For instance, Acme could reasonably consider residential information as relevant for decision making, as companies might legitimately want to hire people who live nearby. Address, however, might act as a proxy for race, which has historically been associated with residential segregation in the United States.

Each of these types of biases is considerably difficult to find, and building models without bias becomes harder as models become more complex[37]. Although there exist certain techniques to reduce a specific type of bias, they are in practice very limited; they work on one definition of bias at a time[38], and typically fail when combined together[39]. Crucially, there is no general consensus on precisely which types of bias are acceptable[40]. What we need instead is transparency into what types of biases exist within a model, and a public-facing discussion about how to best represent our values in the model.

### A.2 Model Structure

Machine learning models have specific algorithmic architectures and use different methods for 'learning' and 'predicting'. For instance, one might use a convolutional neural network for image recognition, but might prefer different architectures for voice parsing. Because the model specification and design fundamentally define how the decision-making process takes place, mistakes in the model structure might lead to unintended consequences. Hence, Acme's algorithms might also be biased because of how their machine learning models are structured. We will now explore the two main scenarios in which this phenomenon can occur.

---

First, the model might be inappropriate if it is being used for something other than the original task it was designed for. For instance, this could occur if Acme used a machine learning model that detects smiles to rate how 'happy' the candidate is. Yet recent literature suggests that the intuitive correlation between the emotion and facial expression is unfounded, as expressions vary significantly across different communities and cultures, and as feelings encompass a much broader range of expressions[41]. This application of models to other areas is not uncommon: despite the growing success of machine learning across a wide range of applications, the most powerful machine learning models today are heavily specialized for specific tasks[42], and generalizing their scope might often result in biased outcomes[43]. However, by using smiles as a proxy for happiness, Acme is subtly discriminating against cultures with different modes of expression.

Second, the model might simply not be powerful enough to tackle a given task. This could be the case, for example, if Acme uses the current state-of-the-art speech recognition models to transcribe a candidate's speech. Even the best models used by Google and Amazon still fail to properly understand basic accented English[44], even if the accent is American in certain cases[45]. In using these models, Acme is likely discriminating against minorities and foreigners, whose English is less regular, and might more likely be falsely understood as less grammatically correct. We see this also with today's text analysis algorithms, such as Google's Perspective API, which some have claimed is heavily biased against black people[46].

One significant challenge in addressing the potential for failure in model structure is that most machine learning models make decisions that are hard or impossible to interpret and explain. Thus in many cases, the designers of the algorithms themselves are not aware of the biases within their models. By requiring transparency, however, not only will the algorithm designers be strongly incentivised to ensure that their algorithms are not obviously biased, but, as we later argue, civil society is able to engage in a dialogue with the creators.

### A.3 Metrics

Besides the structure of the model and the dataset used to train it, we also care about how it actually performs. We use 'metrics' broadly to refer to the set of statistics which assess the model's performance.

Generally, the quality of a model is evaluated based on different error metrics (e.g, false positive rates and false negative rates are used for measuring performance error). A machine learning model might be considered good enough to be applied in social settings if some of those metrics are shown to be within a given margin, or if some of those metrics are equal across different classes of people. For example, Acme might require that the number of false negative rates are equal for white people and black people. This means that if the algorithm wrongly decides that a person is not fit for the job, it must do so with the same frequency for black and white people alike. In mandating that black and white people be treated the same, Acme is formalizing fairness. More generally, we can define fairness as a metric requirement used by algorithm designers to determine equity and equality in decision-making algorithms[47].

Like bias, however, fairness has proven to be a tricky concept in the field of machine learning. Fairness has been described as an unstandarized, subjective concept. As mentioned in part A.2, there is no consensus on the definition of fairness. However, by formalizing a certain definition of fairness, algorithm designers grant an illusion of objectivity, resulting in what Selbst et al.[48] call the "formalism trap". The complex sociopolitical and historical landscape is extremely difficult to formalize in the language of mathematics, and what we often observe instead are algorithms that ignore important social facts.

---

[47] Cornell University. "Are hiring algorithms fair? They're too opaque to tell, study finds." ScienceDaily. ScienceDaily, 20 November 2019. <www.sciencedaily.com/releases/2019/11/191120175616.htm>.
[48] Selbst, Andrew D. and Boyd, Danah and Friedler, Sorelle and Venkatasubramanian, Suresh and Vertesi, Janet, Fairness and Abstraction in Sociotechnical Systems (August 23, 2018). ACM Conference on Fairness, Accountability, and Transparency (FAT*), Vol. 1, No. 1, Forthcoming . Available at SSRN: https://ssrn.com/abstract=3265913

Recent literature suggests that most of today's common definition of fairness can, in fact, reinforce unfair results[49]. For instance, *calibration fairness* is the requirement that, for the same set of considered characteristics, different groups are treated equally by the machine learning model. In other words, that prediction results are independent of protected characteristics. In this scenario, if an employer uses residential location as input to a model, black and white people living in the same residential area should be scored the same. However, as we have seen, residential location and race are strongly correlated in the United States[50]. Hence, requiring that this specific definition of fairness means enforcing the algorithm to generate discriminatory results.

This single example illustrates how a seemingly intuitive definition of fairness might still reproduce bias. In fact, in their review of fairness in machine learning, Sam Cobertt-Davis and Sharad Goel contend[51] that the most commonly used definitions for fairness are insufficient for establishing equity, and in some cases might reinforce existing biases. The question of which metrics we should prioritise is very complex, and we do not attempt to address it within this paper. Instead, we claim that these critical decisions— such as Acme's choice to require false negatives are equal for blacks and whites— need to be transparent and hiring software companies need to be held accountable for them.

### A.4 Applications

Another way in which the model may fail is through being applied in a way that justifies existing biases in the hiring system. The risk assessment algorithms used in many U.S. Courts[52] provide an excellent example. When these algorithms were introduced into the Judicial system, we observed a range of responses from the judges. Depending on the jurisdiction, judges were allowed to use the risk scores in a "constrained or discretionary" means[53]. Moreover, there are suggestions that judges treat the scores differently after being exposed to them more frequently[54]. This should be a significant concern because *it may void all prior attempts at fairness*. Even if the risk assessment tool is perfectly calibrated and fair, if a judge decides to use the tool differently for young versus old offenders, then the outcome is not calibrated nor fair.

---

[49] Corbett-Davies, S., & Goel, S. (2018). The Measure and Mismeasure of Fairness: A Critical Review of Fair Machine Learning. ArXiv, abs/1808.00023.

[50] Osypuk et al. Quantifying Separate and Unequal: Racial-Ethnic Distributions of Neighborhood Poverty in Metropolitan America (2010).

[51] See [49].

[52] See https://epic.org/algorithmic-transparency/crim-justice/ for more information.

[53] See note 43.; State v. Loomis, 881 N.W.2d 749, 774 (Wis. 2016)

In the case of Acme, even if their AcmeSuite tool were entirely fair and unbiased, if the hiring manager decides to use the scores to dismiss the candidates he or she does not like, but ignores the scores when he or she likes the candidates, then the tool is most likely going to reinforce existing prejudices.

A related concern is that the tool is used to justify existing power structures. Computer scientist Rob Kling terms this recurring phenomenon 'reinforcement politics'[55]. Kling found that the top officials in many small cities used new computer systems to "tighten their control over departments and to gain power at the expense of the part-time city councils". Others have pointed to how the introduction of the CT scanner in hospitals led to a power struggle between radiologists and technicians[56]. This is problematic because it may also serve to entrench the biases of those in power. If the existing hiring manager uses AcmeSuite to consolidate power at the expense of less experienced employees, it will be that manager's biases that have the most influence over the hiring process. Yet this information is extremely difficult to procure; there are no laws mandating the transparency of tool use in hiring. Instead, we are left hoping that these hiring algorithms are trained with representative data, powerful enough to make the inferences they claim to be able to do, and are used properly.

As the examples raised in Section II.C and in this section have demonstrated, we cannot simply rely on the promises provided by hiring software companies or their customers. Besides the fact that they may not even be aware of the biases within their own systems, they are incentivized to build a product that meets the minimal standards for 'fair'. Yet when the meaning of 'fair' is in dispute, and the scientific community's understanding of these systems is evolving, it seems very unlikely that these hiring software companies happen to have the perfect solution to these thorny issues around fairness. Civil advocate groups and research groups need to be able to see the outline behind how these tools were designed, trained, built and used in order to ensure that fairness in theory equates to fairness in practice. However, as we discuss in the next section, the current legal landscape is ill-suited to mandating transparency.

---

[54] Stevenson, Megan. "Assessing risk assessment in action." *Minn. L. Rev.* 103 (2018): 303.
[55] Kling, Rob. "Computerization and social transformations." *Science, Technology, & Human Values* 16.3 (1991): 342-367.
[56] Stephen R. Barley. 1996. Technology as an occasion for structuring: evidence from observation of CT scanners and the social order of radiology departments. Administrative Science Quarterly 31 (1996), 78–108.; also see note 43.

IV.   **The Law**

Within this section, we begin by surveying the background of employment discrimination law. We then argue that current employment discrimination law is severely hindered by the 'transparency gap' in hiring software, making it ill-suited for the recent advances in algorithmic hiring tools. Finally, we discuss the role of trade secret law in allowing both employers and hiring software companies to evade responsibility.

A.   **Existing Employment Law**

In the U.S., the Equal Employment Opportunity Commission (EEOC) is the primary regulatory body[57] responsible for preventing discrimination in the process of hiring. It was established in 1964 by the Civil Rights Act, which itself mandated in Title VII that sensitive attributes (race, color, religion, sex, and national origin) cannot serve as a basis for employment discrimination[58].

The most relevant document outlining the standards for employment is the Uniform Guidelines on Employment Selection Processes[59] (1978). There are two significant frameworks introduced by the Uniform Guidelines that are still in effect today. The first establishes the validity of a pre-employment assessment, whilst the second provides the criteria necessary to challenge the decision.

According to Title VII of the Uniform Guidelines, a pre-employment assessment is required to be *valid*. In order for the assessment to be *valid*, it must obey one of three types of validity: *criterion*, *content*, or *construct*. *Criterion validity* refers to whether or not the assessment is predictive of job performance. *Content validity* asks the question 'is this assessment directly relevant for the job that the candidate will be performing?' *Construct validity* refers to the notion that the assessment is testing an underlying characteristic (e.g. leadership) which is relevant to the job. If each of the elements of an assessment obeys at least one of the three criteria, the assessment can be considered valid.

---

[57] While the EEOC makes the rules, the U.S. Department of Justice is responsible for enforcing discrimination cases.
[58] *EEOC Home Page*, https://www.eeoc.gov/eeoc/history/35th/thelaw/civil_rights_act.html.
[59] "Uniform Guidelines." *Uniform Employee Selection Guidelines on Employee Selection Procedures*, http://www.uniformguidelines.com/uniform-guidelines.html.

In order to accuse an assessment of being discriminatory, the Uniform Guidelines stipulate that the plaintiff needs to demonstrate either *disparate treatment* or *disparate impact*. *Disparate treatment* can be simply shown by demonstrating that the assessment explicitly uses a protected characteristic such as race or sex, thus violating Title VII under the Civil Rights Act. *Disparate impact* is a more nuanced criterion[60]. Roughly speaking, it establishes what is known as the '⅘ rule'. This heuristic holds that if the selection rate for a protected group is less than ⅘ of the selection rate of the group with the highest selection rate, the employer may be hiring using discriminatory practices. An employer is able to defend itself by arguing that its assessments were valid and necessary from a business perspective. However, if the plaintiff is able to provide an alternate selection procedure of comparable cost with less adverse impact, the employer may still be liable[61].

### B. Title VII Fails to Address Algorithmic Hiring Discrimination

Under Title VII, along with the Uniform Guidelines that together comprise today's employment law, it is very difficult to demonstrate algorithmic hiring discrimination.

Existing judicial cases regarding hiring discrimination tend to favor employers over job candidates. Two significant Supreme Court rulings provide strong evidence. In *Albemarle Paper Co. v. Moody*, the Court ruled that an employer can avoid claims of disparate impact under Title VII by proving that his or her screening tests are "job related … and consistent with business necessity."[62] Further, *Wards Cove Packing Co. v. Atonio*[63] demonstrated that simply examining the demographic statistics of job positions is inadequate to demonstrate hiring discrimination, because the statistics may reflect the racial disparity in the labor market at the time. As a result, in order to demonstrate disparate impact, there is a significant burden of proof placed on the job candidate. Empirically, we can also see that employees have had little excess filing and winning a disparate impact claim. Legal scholar Michael Selmi found that plaintiffs of disparate impact cases only had a 19.2% success rate from 1984-2002 for Court of Appeals decisions, and a 25.1% success rate from 1983-2002 for District Court decisions[64]. Thus, a large majority of disparate impact claims from both courts are ruled in favor of the employer. There are very few successful disparate impact cases outside of the context in which the doctrine was developed.

---

[60] Raghavan, Manish, et al. "Mitigating Bias in Algorithmic Employment Screening: Evaluating Claims and Practices." *SSRN Electronic Journal*, 2019, doi:10.2139/ssrn.3408010.
[61] Ibid. (see 38)
[62] "*Albemarle Paper Co. v. Moody*, 422 U.S. 405 (1975)." *Justia Law*, https://supreme.justia.com/cases/federal/us/422/405/.
[63] "*Wards Cove Packing Co. v. Atonio*, 490 U.S. 642 (1989)."

*Albemarle Paper Co. v. Moody* also provides an excellent example of how existing employment law is problematic when applied to algorithms. While employers can ignore disparate impact claims by demonstrating that their selection criteria is "job-related", since job-relatedness is typically decided by a statistical correlation, as others have pointed out[65], and algorithms select factors by their statistical correlation, of course the algorithms will be able to claim that their criteria are "job-related". To continue the example raised in the previous section, if Acme Hiring Corporation were to be brought to court, even if their algorithms were biased against a subgroup (e.g. older applicants), they would likely be allowed to continue to sell their software as long as they argue that their algorithms improve production, efficiency, and employee retention for their customers[66].
The extent of Title VII's inadequacies when applied to algorithms becomes more apparent when we consider the criterion of disparate treatment.

As mentioned in Section IV.A, *disparate treatment* criterion requires that sensitive characteristics such as race or gender must not be taken into account when hiring. But because algorithmic models independently and dynamically *learn* the relationships between possibly discriminatory variables, hiring algorithms lack many of the qualifications necessary for making a disparate impact claim. Though there may be no intentional bias and no *human-produced* disparate impact, there are many cases in which a hiring algorithm could associate a protected demographic group (race, ZIP code, physical appearance) with

---

a lack of "employability", voiding Title VII's attempt at regulating disparate impact. Unfortunately, given the aforementioned difficulty in proving a disparate impact case and the fact that Title VII was not created to stop biased *algorithms*, this discriminatory correlation would *not* be prohibited under Title VII law.

We also observe a limitation in the U.S. Department of Justice's practical ability to enforce cases concerning algorithmic decision making. In *Loomis v. Wisconsin*, a Supreme Court case dealing with the closed-source risk-assessment system of COMPAS[67], the court demonstrated a fundamental misunderstanding of how COMPAS operates. As legal scholar Leah Wisser has pointed out[68], the court had interpreted COMPAS as an individual risk assessment metric, yet the algorithm itself is designed to distinguish between the risk levels of "offenders who have similar characteristics". For example, a man with a prior violent convictions may receive a low score of 4, indicating that "60% of the population looks more risky in that area than [the offender] does, and 30% looks less risky"[69] while a woman with no prior offenses may also receive a score of 4, yet be far less likely to commit a crime again. This bears emphasis. In the most significant ruling to date involving algorithmic decision making, the Supreme Court of Wisconsin misunderstood how the algorithm in question worked[70]. Yet it is not entirely their fault. Wisser points out later that the whole section on understanding the COMPAS score was removed from their Practitioner's Guide between 2012 and 2017[71]. The lack of transparency around algorithmic decision making has already had negative consequences in court cases.

### C. Trade Secret Law Enables The Transparency Gap

As legal scholars have pointed out[72], companies have turned to trade secret law— originally intended to protect valuable business information from espionage— to conceal information from civil advocacy groups and consumers. The Defend Trade Secrets Act

---

[67] 881 N.W.2d 749 (Wis. 2016)

[68] Wisser, Leah. "Pandora's Algorithmic Black Box: The Challenges of Using Algorithmic Risk Assessments in Sentencing." *Am. Crim. L. Rev.* 56 (2019): 1811.

[69] Practitioner's Guide To COMPAS, NORTHPOINTE 5 (Aug. 17, 2012), http://www.northpointeinc.com/files/technical_documents/FieldGuide2_081412.pdf[hereinafter Practitioner's Guide 2012].

[70] State v. Loomis, 881 N.W.2d 749, 774 (Wis. 2016) (Abrahamson, J., concurring) (noting the "court's lack of understanding of COMPAS was a significant problem").

[71] Compare Practitioner's Guide 2017, and Practitioner's Guide to COMPAS, NORTHPOINTE (Mar. 19, 2015), http://www.northpointeinc.com/downloads/compas/Practitioners-Guide-COMPAS-Core-_031915.pdf, with Practitioner's Guide 2012.

[72] Deepa Varadarajan, Trade Secret Fair Use, 83 Fordham L. Rev. 1401 (2014), (see note 61 below)

(DTSA), in particular, federalized trade secrets laws, giving companies much more protection in the realm of trade secrets. Specifically, trade secrets must 1) not be generally known, 2) bring economic value to its holder by virtue of it not being known, and 3) be subject to reasonable precautions to keep the information secret[73]. In practice this means that software companies can use the DTSA as an excuse to withhold information. In one public example, Palantir Technologies, which had provided the New York Police Department (NYPD) with information derived from police data (e.g. broader trends in crime), refused to hand over an unencrypted version of its information to the NYPD when the department decided to partner with IBM instead[74]. To the best of our knowledge, there have not been any court cases against either employers or hiring software companies for algorithmic hiring discrimination, we fully expect that the Data Trade Secrets Act will be invoked by hiring software companies as a shield to scrutiny from both the Department of Justice and the public.

### D. Civil Society is Powerful, if Informed

As Economist Daron Acemoglu argues in his book, *Why Nations Fail*, democracies' social stability rely in large part on the strength of civil society. He tells the story of J.P. Morgan, one of the 'robber barons' in early twentieth century America, decreasing prices until competitors went bankrupt, then buying the competitors, contributing in large part to wealth inequality. Only through the efforts of individual 'muckrakers' such as Louis Brandeis, in publishing his famous book *Other People's Money and How The Bankers Use It*[75], did public awareness increase enough to create the Pujo Committee, which investigated the financial 'money trust', and other financial regulations.

Civil society has been highly involved and effective at influencing technology policy to promote the rights of consumers. In 1997, the American Civil Liberties Union brought a case against the Attorney General of the United States to the Supreme Court arguing that the anti-indecency provision of the 1996 Communications Decency Act violated the First Amendment's guarantee of freedom of speech[76]. The result was a unanimous ruling that the Internet is entitled to the full protections given to media like the printing press. This landmark decision set a strong precedent for protecting free speech online and, as the Electronic Frontier Foundation Senior Counsel David Sobel remarked, "defined the First Amendment for the 21st century … [establishing] the fundamental principles that govern free speech issues in the electronic age."[77] We also observed a significant shift in public attention with ProPublica's reporting on the use of the COMPAS algorithm for recidivism rate prediction in U.S. Courts[78]— at least 578 scholarly articles cited either the ProPublica "Machine Bias" article between May 2016 and December 2017[79]. In arguing that the COMPAS algorithm had higher false positive rates for black defendants than white defendants, they not only ignited significant debate over the issue of algorithmic fairness, but provided a new methodology for others to argue about algorithms[80].

---

[73] See 18 U.S.C. § 1839(3)(B) (2018); Metallurgical Indus., Inc. v. Fourtek, Inc., 790 F.2d 1195, 1199 (5th Cir. 1986).
[74] Alden, William. "There's A Fight Brewing Between The NYPD And Silicon Valley." *BuzzFeed News*, BuzzFeed News, 28 June 2017, https://www.buzzfeednews.com/article/williamalden/theres-a-fight-brewing-between-the-nypd-and-silicon-valley; "Palantir Contract Dispute Exposes NYPD's Lack of Transparency." *Brennan Center for Justice*, https://www.brennancenter.org/our-work/analysis-opinion/palantir-contract-dispute-exposes-nypds-lack-tr ansparency.
[75] Brandeis, Louis Dembitz. *Other Peoples Money: and How the Bankers Use It*. Martino Publishing, 2009.

ProPublica were only able to do this because, as they note, "Florida has strong open-records laws,"[81] which allowed them to submit a public records request for all 18,610 people assessed between 2013 and 2014.

Similarly, in 2016, they were able to demonstrate that Facebook advertising enabled exclusion based on race, age, and gender. This caused a set of lawsuits from civil rights organizations against Facebook that ultimately resulted in advertisers no longer being able to segregate among their advertising audience on Facebook[82]. This was only possible because ProPublica could pose as an advertiser and collect the information leading to its findings.

The Electronic Privacy Information Center (EPIC), a nonprofit concerned with protecting digital privacy, provides yet another example. In 2013, EPIC filed a complaint with the FTC against Snapchat[83], the photo sharing app, for deceptive business practices. Snapchat claimed that its photos would "disappear forever", however, in reality others could access photos even after users had been notified of the photos' deletion. As a result, the FTC filed a 20-year Consent Order. Snapchat is now subject to 20 years of privacy audits, and is prohibited from making false claims about its privacy practices[84].

---

[76] *Reno v. American Civil Liberties Union*, 521 U.S. 844 (1997)
[77] "Ten Years After ACLU v. Reno: Free Speech Still Needs Defending." *Electronic Frontier Foundation*, 9 Feb. 2012, https://www.eff.org/effector/20/25.
[78] Larson, Jeff, et al. "How we analyzed the COMPAS recidivism algorithm." *ProPublica (5 2016)* 9 (2016).
[79] Washington, Anne, How to Argue with an Algorithm: Lessons from the COMPAS ProPublica Debate (February 4, 2019). Accepted for publication. The Colorado Technology Law Journal. Volume 17 Issue 1 http://ctlj.colorado.edu. Available at SSRN: https://ssrn.com/abstract=3357874
[80] Ibid. (see 74)
[81] Ibid. (see 27)
[82] Gillum, Jack, and Ariana Tobin. "Facebook Won't Let Employers, Landlords or Lenders Discriminate in Ads Anymore." *ProPublica*, 10 June 2019, https://www.propublica.org/article/facebook-ads-discrimination-settlement-housing-employment-credit.

We observe repeatedly that civil society is able to effectively champion public interests and actually effect change, *when given the required information.* But what we observe in the field of algorithmic hiring software, despite the significant impact that these algorithms have on candidates' lives, there is little to no transparency regarding how the algorithms are trained, audited for bias, or even how they are used in the hiring process, as established in Section II.C. Anecdotally, when the authors of this paper attempted to get access to a demonstration of HireVue's platform, despite supplying business emails and the other relevant credentials, we were unable to get access to a demonstration.

What we need is a minimal requirement for information about hiring software. With more clarity about the dataset, model structure, metrics, and application of these algorithms, not only will the Department of Justice be provided a clearer understanding of the implications of the technology they are litigating, but civil society will be provided a vital tool to productively effect change.

## V.    Our Solution

At this point in the report, it should be clear that 1) algorithmic bias is a significant problem, 2) four main factors that influence algorithmic bias are the *model structure*, *dataset*, *metrics*, and *application*, 3) these factors can be in large part mitigated by transparency, and 4) there are significant extant legal barriers to ensuring algorithmic fairness. In light of these findings, we propose *algorithmic transparency reports* as an additional EEOC regulation. This would apply to companies that employ machine learning tools for hiring. We argue that these reports will enable civil advocate groups and media

---

outlets to hold employers accountable, and enable healthy public debate regarding the standards of fairness. Later, we address potential concerns about our proposal.

###    A.   Towards More Transparent Algorithmic Use

If, as we have observed, neither the EEOC nor computer scientists are able to establish a specific definition of fairness broad enough to address the desires of different groups[85], then how can we ensure that the decisions made by algorithms are fair?

It is first worth noting that 'fair' depends on context. As an intuitive example, if an employer finds that 'distance from workplace' is a strong predictor of success in the office, can the employer use this as a metric? It depends. If a closer distance from workplace leads to success in the office because people choose to live near their office because they are more passionate about their jobs, then it might be acceptable. If, however, 'distance from workplace' is found to correlate with race, where minority races are found to live *far* from the workplace while the predominant race is found to live *close* to the workplace, then the metric enables indirect racial discrimination. As we argued in Section II.C, attempts to capture 'fairness' in the form of single rule or an equation mostly fail to address the contextual nature of fairness.

Instead, we can acknowledge the contingent nature of fairness by allowing civil society and the press to adjudicate. By civil society, we mean non-profit rights groups such as the Electronic Frontier Foundation, the Electronic Privacy Information Center, and the American Civil Liberties Union, but also investigative journalist groups such as ProPublica.

The use of civil society and the press is appealing because, as we argue in the following section, they have a strong reputation for effecting positive change, especially in technology. In order to do this, however, there needs to be more information available about companies' hiring practices. As established in Section II.C, there have been several individuals and groups that are dissatisfied with the lack of transparency around algorithmic hiring. To remedy the lack of transparency and enable civil advocate groups to better hold hiring software companies accountable, we propose *algorithmic transparency reports*.

---

There has been significant recent work in the fair machine learning community to create summary information regarding the model and the data used to train the model[86]. Two papers that have had a significant influence on our model of *algorithmic transparency reports* are "Datasheets for Datasets"[87] by Gebru et al. and "Model cards for model reporting"[88], by Mitchell et al. "Datasheets for Datasets" introduces the idea of using a 'datasheet' to characterise several aspects of the dataset, including how the data collection process was funded and the coverage of the dataset. 'Model cards' propose a summary sheet about the trained model, ranging from model details (including facts about the specific algorithms used) to the evaluation data that is used to assess the model.

Our proposal differs from both Gebru's and Mitchell's proposals in a few key respects. First, unlike datasheets and model cards, the algorithmic transparency reports are focused specifically on hiring. We have included questions intended to address how the model interacts with U.S. hiring laws such as the American Disability Act[89] and, of course, Title VII of the US Civil Rights Act[90]. Further, our reports aim to document how the model is deployed in practice. As has been pointed out[91], there is often a discrepancy between the intended use of an algorithm, and how it is deployed in reality. For this reason, we believe actual use should be documented, and we have chosen to include it within our framework. A final point of difference between our model and the model cards proposed by Mitchell is that we have explicitly chosen to exclude details of the model in the report to allay concerns regarding intellectual property.

Another framework that bears some similarity to our proposal is the European data protection impact assessment (DPIA)[92]. However, while both our proposal and the DPIA cover the intent and risk factors of data processing, the DPIA is focused specifically on fulfilling the requirements of the European General Data Protection Regulation. Our reports focus on a specific evaluation framework that is intended for both the public and the Department of Justice and is situated firmly in the context of existing U.S. legislation.

---

[86] Bender, Emily M., and Batya Friedman. "Data statements for natural language processing: Toward mitigating system bias and enabling better science." *Transactions of the Association for Computational Linguistics* 6 (2018): 587-604. (see sources 18 and 19 below)

[87] Gebru, Timnit, et al. "Datasheets for datasets." *arXiv preprint arXiv:1803.09010* (2018).

[88] Mitchell, Margaret, et al. "Model cards for model reporting." *Proceedings of the Conference on Fairness, Accountability, and Transparency.* ACM, 2019.

[89] See https://www.ada.gov/

[90] *Title VII of the Civil Rights Act of 1964*, https://www.eeoc.gov/laws/statutes/titlevii.cfm.

[91] Angèle Christin. 2017. Algorithms in practice: Comparing web journalism and criminal justice. Big Data & Society 4, 2 (2017).

[92] "Data Protection Impact Assessments." *ICO*, https://ico.org.uk/for-organisations/guide-to-data-protection/guide-to-the-general-data-protection-regulation-gdpr/accountability-and-governance/data-protection-impact-assessments/.

Directly inspired by the factors that enable bias, the report should contain four broad sections. *Intent, Dataset, Metrics,* and *Applications.*

- **Intent**

    This section should enable a nontechnical audience to grasp what the model was originally intended to do. This should capture many of the concerns raised about model structure in III.A.2. After reading this section, readers should have a good idea of what use cases are within or outside of the intended scope for the model. This section should be supplied by hiring software companies to the employers who use the automated hiring tool. Below are questions to guide the hiring software company in their report of this section.

    - What are the primary uses of the tool?
    - Who are the primary intended users? (Recruiters with no technical background or engineers with technical background?)
    - What is the nature of the tool's output, and how does it relate to job desirability?
    - Could the tool be applied in an inappropriate manner? If so, describe how, including a brief discussion of the pragmatic, legal, and ethical implications.

- **Dataset**

    As highlighted in a recent article[93], even the most widely used dataset contains pernicious value judgments— a child with glasses may be labeled "unsuccessful" or "a loser", and a woman in a bikini "a slut". Further, as we established in Section III.A.1, it is incredibly difficult to 'correct' a model trained on biased data.  It is therefore vital to include summary characteristics of the dataset. This section is intended to reveal the ways in which the dataset may reflect human biases, and the attempts taken to mitigate this.

    - How was the training data collected?

---

[93] Crawford, Kate, and Trevor Paglen. "Excavating AI." -, https://www.excavating.ai/.

- What are the demographics of the people in the training data? Does this deviate from the national averages?
- If there are labels associated with the data, what are the types of labels?
- Who labeled the training data? Could the labels have been affected by the value judgement of the labelers?
- What has been done to validate the labels?
- Was there a test set created to validate the data? If not, why not? Was the test set similarly distributed to the training set?

- **Metrics**

    In order to capture the presence of biases such as the insufficient power of the model, as well as its performance for different groups (discussed in Section III.A.3), it is important to be able to examine the metrics surrounding the data. This section aims to reveal the performance of the model on different protected groups, and to validate attempts to address bias.

    - How was model performance measured, and why were these metrics chosen over other metrics?
    - If thresholds had to be manually chosen, why were those thresholds used?
    - If the model is a classification system, what is the confusion matrix for each of the protected characteristics?
    - If the model is score-based, then how does the distribution of scores differ between people of different protected groups?
    - What measures have been taken to ensure that the algorithms are not biased on the basis of race, color, religion, sex, national origin, or age?
    - Which definition of fairness, if any, have you chosen to use in your effort to mitigate bias?
    - Has the model been tested against people with disabilities, or people that are pregnant?
    - Is there an attempt to measure 'truthfulness' of a candidate's statements, thus potentially violating the Employee Polygraph Protection Act?

- **Applications**

As outlined in Section III.A.4, much potential bias comes from the selective use of model outputs by hiring managers in deciding on candidates[94]. This section aims to make explicit how the model is deployed in practice, and its interaction with other models.

  - Is the model consistently used in practice?
  - How are the outputs of the model used in decision making? As the input to another algorithm, or as a suggestion to a human recruiter?
  - Are there other models being used in conjunction with this model, and if so, could the use of this model violate the intended use of the other models?

In line with calls from several members of Congress[95], we propose that the EEOC Uniform Guidelines be updated to include a section that:

1. *Requires complete and truthful algorithmic transparency reports to be supplied by employers who use algorithmic hiring software to automatically evaluate either employee resumes or video statements to the public in an easily accessible manner;*
2. *Requires that hiring software companies supply employers the required information to fill out the algorithmic transparency report.*

Key Definitions:

A. *Algorithmic transparency report* refers to the questions outlined in the four broad sections described above.
B. *Employer* means a person, as defined by section 701 of the Civil Rights Act of 1964, 42 U.S.C. 2000b, engaged in an industry affecting commerce who has fifteen or more employees.
C. *Algorithmic hiring software* means software intended to, or otherwise used to, automatically assess job applicants.
D. *Hiring software company* means an organization engaged in an industry affecting commerce that produces algorithmic hiring software.

---

[94]Hoffman, Mitchell, Lisa B. Kahn, and Danielle Li. "Discretion in hiring." *The Quarterly Journal of Economics* 133.2 (2017): 765-800.
[95] Gershgorn, Dave. "Senators Are Asking Whether Artificial Intelligence Could Violate U.S. Civil Rights Laws." *Government Executive*, Quartz, 21 Sept. 2018, https://www.govexec.com/technology/2018/09/senators-are-asking-whether-artificial-intelligence-could-violate-us-civil-rights-laws/151475/.

E. *Video statement* means a video recording of a candidate for the use of an employer to assess the candidate for a job.

## B. The Concern of Trade Secrets

As argued in Section IV.C, trade secret law poses a significant barrier to transparency. We, along with many legal scholars, argue that the protections provided by the DTSA are overbroad[96], and hinders vital civic oversight of large companies. However, it is out of the scope of this report to explore in close detail the negative externalities of current trade secret law. Instead, we suggest a slight amendment.

First, we suggest that the DTSA's definition of 'improper means' expand from "does not include reverse engineering, independent derivation, or any other lawful means of acquisition" to "does not include reverse engineering, independent derivation, *technological transparency reports*, or any other lawful means of acquisition." *Technological transparency reports* refer to the superclass of reports that the government may mandate technology companies to provide for the sake of auditing, and includes algorithmic transparency reports. This specific provision accounts allows for greater transparency with no significant difference to the current scope of the DTSA.

We would like to acknowledge that other proposals, such as legal scholar Varadarajan's proposal for a doctrine of 'fair use' within trade secret law[97], would also address the limitations of the trade secret law, and allow for algorithmic transparency reports, and we would welcome similarly inspired reforms. A detailed discussion of trade secret is, however, out of scope for this article.

By requiring that employers release algorithmic transparency reports, the EEOC provides civil society with the necessary information to report and litigate in the interest of the public. This, in turn, creates a strong incentive for employers and hiring software companies to maintain a high standard of algorithmic accountability, lest they be brought to court by civil rights organizations.

It has not escaped our attention that algorithmic transparency reports also have implications with regards to social media screening of candidates, since such a model is likely to be deployed in the middle to late stages of the hiring process. However, since such 'cyber-screening' is also regulated by the Fair Credit Reporting Act[98], its discussion is outside the scope of this article.

---

[96] See note 65, 93, 95
[97] Ibid. See note 65

### C. Anticipating Objections

There are two significant objections that we anticipate, one from the side of the employers and hiring software companies regarding intellectual property violations, and another from the side of the civil rights groups about how a transparency report is inadequate to address the issue of algorithmic bias in hiring.

First, as several legal scholars have pointed out[99], attempts to access data or reverse engineer algorithmic systems are hindered by existing intellectual property laws such as the Digital Millennium Copyright Act (DMCA) and cybersecurity laws such as the Computer Fraud and Abuse Act (CFAA), both of which have been used by software companies to obstruct analysis of their decision making systems[100]. This is a legitimate concern. However, as Professor Sonia Katyal points out[101], copyright law only protects the "precise expression" of an algorithm[102]. In practice, this refers to information such as the exact names of commands that are used. Algorithmic transparency reports do not demand specific details from the model, other than the nature of its inputs and outputs. In this way, the reports may be considered equivalent to asking the driver of a vehicle what type of fuel is being used. The type of fuel allows an observer to discern whether the vehicle is a car, or an airplane, but does not reveal how the car or airplane specifically functions. Further, as recent rulings have suggested, the breadth of the CFAA (the most relevant cybersecurity

---

[98] "Social Media-Based Screening and FCRA." *JD Supra*, https://www.jdsupra.com/legalnews/social-media-based-screening-and-fcra-94902/.

[99] Ajunwa, Ifeoma, Automated Employment Discrimination (March 15, 2019). Available at SSRN: https://ssrn.com/abstract=3437631 or http://dx.doi.org/10.2139/ssrn.3437631 , R. Moore, Taylor, Trade Secrets and Algorithms as Barriers to Social Justice (August, 2017)

[100] "Google avidly protects every aspect of its search technology from disclosure". Nicole Wong, Response to the DoJ Motion, OFFICIAL GOOGLE BLOG (Feb. 17, 2006), https://googleblog.blogspot.com/2006/02/response-to-doj-motion.html, Deepa Varadarajan, Trade Secret Fair Use, 83 Fordham L. Rev. 1401 (2014).
Additionally, EarthCam, Inc. v. OxBlue Corp., No. 15-11893, 2017
WL 3188453, at *9 n.2 (11th Cir. July 27, 2017) and Facebook, Inc. v. Power Ventures, Inc., 844 F.3d 1058 (9th Cir. 2016) were both cases when the CFAA was invoked to argue that competitors in scraping data from a website, violated the law.

[101] Katyal, Sonia K. "Private Accountability in the Age of Artificial Intelligence." *UCLA L. Rev.* 66 (2019): 54.

[102] See COMPAS Licensing Agreement
§ 8.2 (2010), https://epic.org/algorithmic-transparency/crim-justice/EPIC-16-06-23-WIFOIA-201600805-2010InitialContract.pdf [https://perma.cc/6Y6H-XN9M]; Katyal, *supra* note 34

law to our proposal) does not extend to prohibit auditing and algorithmic analysis by researchers[103]. We believe neither copyright law nor cybersecurity law can be reasonably invoked to prevent transparency.

Instead, as we addressed in Section IV.C, companies have turned to trade secret law— originally intended to protect valuable business information from espionage— to conceal information from civil advocacy groups and consumers. We believe that an amendment to trade secret law, either in the form that we suggested, or within the more comprehensive framework that scholars such as Deepa Varadarajan suggest[104], would maintain the legitimate interests of hiring software companies to maintain trade secrets while still allowing for some degree of transparency.

Another potential objection comes from the side of the public interest. We have observed widespread support for a ban on the use of facial recognition in public areas[105], and some may argue that we should similarly ban the use of facial recognition and other machine learning applications in hiring until we have a broader consensus on how they should be used. However, we feel that this neglects the significant potential benefit to society that automated hiring software can bring. Aside from the performance of today's algorithms, however, it is clear that today's human decisions suffer from significant bias. Whether it is a bias towards physically attractive people[106], a racial bias[107], or even the tendency to underestimate one's own bias[108], humans systematically make decisions that unfairly privilege one group over another. Algorithms, while suffering from biases of their own, have the potential to reduce many of these biases, and we believe that pursuing fair hiring algorithms are part of the answer in creating a just and equitable society.

To continue the example we began in Section III, if Acme Hiring Corporation had been required to release an algorithmic transparency report, not only would the racial bias have likely been caught, but the public dialogue around algorithmic fairness would have been advanced. Whether it is through the Department of Justice pursuing claims of hiring discrimination, or through a news organization like ProPublica pointing to a category that was neglected in the dataset, Acme Hiring Corporation is forced to now take part in the dialogue about algorithmic fairness.

---

[103] Sandvig v. Sessions, 315 F. Supp. 3d 1 (D.D.C. Mar. 30, 2018).
[104] Ibid. (see 38)
[105] Brandom, Russell. "Bernie Sanders Calls for a Ban on Police Use of Facial Recognition." *The Verge*, The Verge, 19 Aug. 2019, https://www.theverge.com/2019/8/19/20812032/bernie-sanders-facial-recognition-police-ban-surveillance-reform?campaign=SkimbitLtd&ad_group=66960X1514734Xad4877c146aadd41fc6285fe499b6321&keyword=660149026&source=hp_affiliate&medium=affiliate.
[106] Shahani-Denning, Comila. "Physical attractiveness bias in hiring: What is beautiful is good." *Hofstra Horizon* (2003): 14-17.
[107] Sporer, Siegfried Ludwig. "The cross-race effect: Beyond recognition of faces in the laboratory." *Psychology, Public Policy, and Law* 7.1 (2001): 170.
[108] Pronin, Emily. "Perception and misperception of bias in human judgment." *Trends in cognitive sciences* 11.1 (2007): 37-43.

**VI.     Conclusion**

There is little doubt that the power of algorithmic systems will continue to grow, yet it is unclear whether the growth will be beneficial for all parties. We have described the legal and technical challenges surrounding algorithmic hiring systems, and have proposed regulations to harness the strength of civil society to address these challenges. We have seen the promise of these algorithms. However, if we are not careful, there is a very real possibility that we observe the erosion of our civil rights. We need to be able to have critical, public-facing discussions about the construction of these systems. This is the only way to maintain equity in an age of hyper-digital connectivity.